\documentclass[conference, 10pt, a4paper]{IEEEtran}
\IEEEoverridecommandlockouts
\usepackage{cite}
\usepackage{amsmath,amssymb,amsfonts}
\usepackage{algorithmic}
\usepackage{graphicx}
\usepackage{textcomp}
\usepackage{xcolor}
\def\BibTeX{{\rm B\kern-.05em{\sc i\kern-.025em b}\kern-.08em
    T\kern-.1667em\lower.7ex\hbox{E}\kern-.125emX}}

\usepackage{wrapfig}
\usepackage{caption}
\usepackage{subcaption}
\usepackage{booktabs}
\usepackage{url}
\usepackage{cleveref}
\usepackage{listings}

\lstdefinelanguage{JavaScript}{
  keywords={typeof, new, true, false, catch, function, return, null, catch, switch, var, if, in, while, do, else, case, break},
  keywordstyle=\color{blue}\bfseries,
  ndkeywords={class, export, boolean, throw, implements, import, this},
  ndkeywordstyle=\color{darkgray}\bfseries,
  identifierstyle=\color{black},
  sensitive=false,
  comment=[l]{//},
  morecomment=[s]{/*}{*/},
  commentstyle=\color{purple}\ttfamily,
  stringstyle=\color{red}\ttfamily,
  morestring=[b]',
  morestring=[b]"
}

\graphicspath{{img/}}

\begin{document}
\title{CryptoScratch: Developing and evaluating a block-based programming tool for teaching K-12 cryptography education using Scratch}
\author{\IEEEauthorblockN{Nathan Percival\IEEEauthorrefmark{1}\IEEEauthorrefmark{3}, Pranathi Rayavaram\IEEEauthorrefmark{1}, Sashank Narain\IEEEauthorrefmark{1} and Claire Seungeun Lee\IEEEauthorrefmark{2}}
\IEEEauthorblockA{\IEEEauthorrefmark{1}Department of Computer Science, University of Massachusetts Lowell, Lowell, MA, USA\\
}
\IEEEauthorblockA{\IEEEauthorrefmark{2}School of Criminology \& Justice Studies, University of Massachusetts Lowell, Lowell, MA, USA}
\IEEEauthorblockA{\IEEEauthorrefmark{3}Middlesex Community College, Lowell, MA, USA\\
nathan\_percival@student.uml.edu, nagapranathi\_rayavaram@student.uml.edu,
sashank\_narain@uml.edu, claire\_lee@uml.edu
}}


\maketitle
\begin{abstract}
The world continues to experience a shortage of skilled cybersecurity personnel. The widely accepted solution for reducing this gap is raising awareness about cybersecurity. In response, many schools are integrating cybersecurity into the K-12 curriculum. Also, educational initiatives from the National Initiative for Cybersecurity Education, GenCyber, and the CryptoClub project enable universities to provide enrichment to middle- and high-school students regarding the importance of cybersecurity. Unfortunately, there is currently an absence of visual and straightforward tools that limits the feasibility of hands-on practice during these initiatives.
This paper presents the design, implementation, and evaluation of a new framework called CryptoScratch, which extends the Scratch programming environment with modern cryptographic algorithms (e.g., AES, RSA, SHA-256) implemented as visual blocks. Using the simple interface of CryptoScratch, K-12 {students} can study how to use cryptographic algorithms for services like confidentiality, authentication, and integrity protection; and then use these blocks to build complex modern cryptographic schemes (e.g., Pretty Good Privacy, Digital Signatures). In {addition}, we present the design and implementation of a Task Block that provides students instruction on various cryptography problems
and verifies that they have successfully completed the problem. The task block also generates feedback, nudging learners to implement more secure solutions for cryptographic problems. An initial usability study was performed with 16 middle-school students where students were taught basic cryptographic concepts and then asked to complete 
tasks using those concepts.
Once students had knowledge of a variety of basic cryptographic algorithms, they were asked to use those algorithms to implement complex cryptographic schemes such as Pretty Good Privacy and Digital Signatures.
Using the successful implementation of the cryptographic and task blocks in Scratch, the initial testing indicated that $\approx$~60\% of the students could quickly grasp and implement complex cryptography concepts using CryptoScratch, while $\approx$~90\% showed comfort with cryptography concepts and use-cases.
Based on the positive results from the initial testing, a larger study of students is being developed to investigate the effectiveness across the socioeconomic spectrum.
\end{abstract}


\section{Introduction}

In today's data-driven world, cybersecurity is crucial for data protection. Many government and private organizations are vulnerable to threats from external actors, and the quality and quantity of cybercrimes keep getting more sophisticated every year. One example of a sophisticated attack is a ransomware attack that surged to record levels in 2021~\cite{Isaca_ransomware}. A disturbing trend of cyberwars is also enfolding between nations resulting in many new disruptive and devastating attack strategies. To defend against such sophisticated attacks, we need a highly qualified cybersecurity workforce. Unfortunately, according to surveys and interviews conducted worldwide, there is a vast and growing cybersecurity talent shortage. For instance, as per a cybersecurity workforce study by (ISC)$^2$, the US has a shortage of 359,236 cybersecurity professionals and such gaps are common across the world with a global requirement of about 3.12 million professionals~\cite{isc2_cybersecurity_2020}. According to ISACA, 61\% of companies feel that they are somewhat or significantly understaffed in cybersecurity and 55\% currently have unfilled cybersecurity positions~\cite{isaca_state_2021}.

In order to address this deficit, government, academia and industry professionals have all stepped up and introduced many exciting and encouraging programs to bring awareness and develop an interest in cybersecurity among children from diverse backgrounds  (e.g., GenCyber~\cite{gencyber_gencyber_2020}, CyberPatriots~\cite{northrop_grumman_foundation_air_nodate}, NICE~\cite{nist_national_2016}, PLTW~\cite{pltw_pltw_2021}). We believe that these programs are and will keep encouraging many students to take up cybersecurity as their profession. Studies in the past also back our belief by showing evidence that increasing awareness of cybersecurity among K-12 students has raised their interest in learning more about cybersecurity career paths~\cite{cyberorg_state_2020}. In addition to these initiatives, Computer Science curricular developments across schools and universities have been adding cybersecurity as a necessary component~\cite{joint_task_force_on_cybersecurity_education_cybersecurity_2017, sabin_acmieee-cs_2016, rowe_role_2011} to their curriculum in addition to creating camps (e.g., GenCyber) to teach cybersecurity to K-12 students \cite{gencyber_gencyber_2020}.

\begin{figure*}[t]
  \centering
  \includegraphics[width=.7\textwidth]{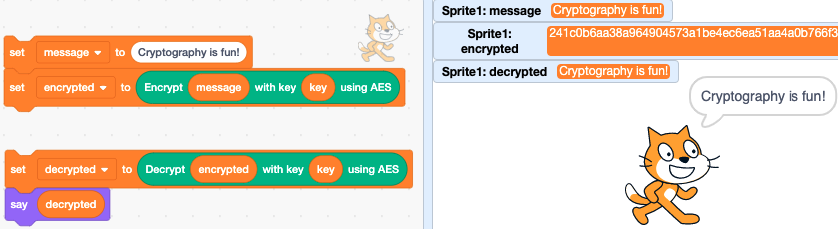}
  \caption{AES Encryption and Decryption using CryptoScratch showing Scratch output dialog}
  \label{fig:CryptoScratchAESBlock}
  \vspace{-1em}
\end{figure*}

Even after updates to the K-12 curriculum and the introduction of camps, only 5\% of high-school students and 1\% of middle-school students in the US are aware of topics such as Cryptography~\cite{cyberorg_state_2020}. Cybersecurity education is still highly underrepresented in middle-school curriculum~\cite{kurt_peker_developing_2020}. As Cryptography is an integral part of cybersecurity, systems and projects such as CryptoClub \cite{university_of_chicagos_center_for_stem_education_cryptoclub_2016}, programs run by the National Cryptologic Museum \cite{nsa_national_2020} and the K-12 Classroom Support from NSA \cite{nsa_request_2021} have created resources to effectively expose students to cryptography and encourage them to explore cryptographic algorithms. Unfortunately, due to the absence of tools for hands-on practice, these programs are often limited to simple ciphers such as the Caesar cipher. We believe that the intuition behind teaching the Caesar cipher is its simplicity and its ability to be computed by hand or using simple tools. Teaching and practicing modern cryptographic algorithms like AES and RSA, on the other hand, requires the knowledge of complex tools or programming languages such as Python that may not be intuitive for young children. In addition, these algorithms have mathematical premises (e.g., finite fields, modular arithmetic) and are often taught to students using these premises, which is not feasible in the K-12 curriculum.


This paper focuses on two related research questions: \emph{(RQ1) can we create a simple, intuitive, and visual block-based tool for K-12 students to learn and practice modern cryptographic algorithms without delving into the mathematical complexities of these algorithms?} and \emph{(RQ2) how well can K-12 students understand and implement complex cryptographic schemes that drive modern systems and applications (e.g., PKI, Digital Signatures)?}  We note that RQ1 focuses on implementing a practical hands-on tool compatible with the systems mentioned earlier, and RQ2 delves into understanding how the usage of this tool can impact the computation thinking of students and the learning of cryptographic algorithms and schemes.

\begin{figure}[t]
  \centering
  \includegraphics[width=0.95\columnwidth]{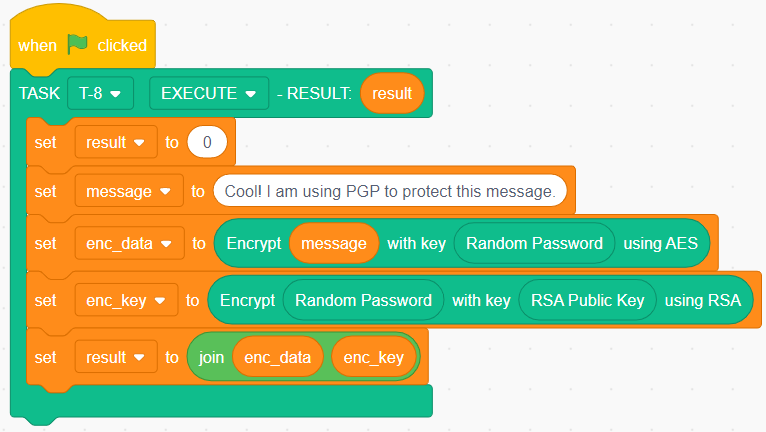}
  \caption{Example implementation of Pretty Good Privacy (PGP) using CryptoScratch Task Block}
  \label{fig:CryptoScratchPGPCode}
\end{figure}

To answer RQ1, we have implemented a novel framework called CryptoScratch as an extension of the  Scratch~\cite{scratch_foundation_scratch_nodate} programming language. CryptoScratch implements modern cryptographic algorithms such as AES, RSA, and SHA-256 as visual blocks. These blocks abstract away the mathematical details of the algorithms to enable students to interact with these blocks easily and learn cryptography without learning the mathematical premises behind the algorithms. \Cref{fig:CryptoScratchAESBlock} shows an example of the implementation of symmetric cryptography visually using the AES Blocks in CryptoScratch. The CryptoScratch framework also implements a Task Block that performs two functions to nudge students towards writing more secure cryptographic schemes. First, it enables educators to build new problems/challenges that students can solve to improve their cryptography knowledge. Second, it enables students to get feedback about their solution’s security and accuracy, enabling them to try different cryptographic schemes and analyze their outputs. \Cref{fig:CryptoScratchPGPCode} shows an example of a task block that checks a student's solution for accuracy of the Pretty Good Privacy (PGP) scheme. We note that the CryptoScratch framework has been implemented as a Scratch extension and will be made open-source and publicly available through the MIT Scratch website. Given its simplicity and wide availability, we also emphasize that the framework can be easily integrated into the K-12 curriculum and the aforementioned programs (e.g., CryptoClub, GenCyber) to enable students to practice the cryptographic concepts taught in their classes and in these programs.

To answer RQ2, we performed a preliminary evaluation of CryptoScratch through a 3-day workshop comprising 16 middle-school students. The goal was to understand how well students can understand cryptography, understand different security services, and build complex cryptographic schemes using CryptoScratch.  The workshop taught these students several concepts such as symmetric and asymmetric cryptography, hashing algorithms, services such as confidentiality, integrity protection, and authentication, and modern schemes such as PGP and Digital Signatures. We also gave the students eight challenges and analyzed their solutions. Based on a pre-post survey conducted with the 16 students and their challenge solutions, we observed that $\approx~60\%$ of the students could easily understand and implement the above complex schemes using CryptoScratch ($\approx~90~-~100\%$ for simpler challenges). Meanwhile, $\approx 90\%$ showed increased comfort with cryptography after the end of the workshop. These numbers indicate that CryptoScratch can be employed as an effective tool to teach cryptography in K-12 classrooms worldwide.

\vspace{.5em}
\noindent In summary, we make the following contributions:
\begin{itemize}
    \item We have designed and implemented a new framework called CryptoScratch as a Scratch extension that implements a selection of modern cryptographic algorithms such as AES, RSA, and SHA-256 as visual blocks on the Scratch platform.
    \item We have designed and implemented a Task Block on CryptoScratch that analyzes students' solutions for security problems and accuracy, and nudges them towards more secure solutions.
    \item We have implemented eight cryptographic learning challenges using the new Task Block that students can use to learn different aspects of cryptography such as symmetric cryptography, asymmetric cryptography, hashing algorithms, and combining cryptosystems.
    \item We evaluated the usability and effectiveness of CryptoScratch through a workshop focused on teaching cryptography using CryptoScratch to 16 middle-school students. We also conducted pre- and post-test user surveys. Approximately $60\%$ of students showed comfort implementing complex cryptographic schemes, while $\approx 90\%$ showed comfort with cryptography concepts.
\end{itemize}

The rest of this paper is structured as follows. In \Cref{sec:relatedwork}, we discuss some of the related works that focus on cybersecurity and cryptography education with K-12 students and compare our CryptoScratch framework with these prior works. \Cref{sec:design} focuses on the design and implementation of the CryptoScratch framework, the design of the Task Block, and the description of cryptographic challenges implemented in CryptoScratch. \Cref{sec:userstudy} describes the design, the metrics, and the results of the user study that was conducted with 16 middle-school students to evaluate the effectiveness of CryptoScratch. We conclude and discuss future work in \Cref{sec:conclusion}.

\section{Related Work}\label{sec:relatedwork}


We acknowledge that many initiatives exist that focus on cybersecurity education and awareness for diverse socioeconomic groups, education of women and other underrepresented groups~\cite{codebreakers_2020,girlsgocyberstartwebsite,womenscyberjutsu_2020,NCL_2020}. This section highlights only a subset of these initiatives. Our focus is on demonstrating how these initiatives encourage cybersecurity education and how CryptoScratch can enable the hands-on practice of cryptographic algorithms and schemes during these initiatives. GenCyber \cite{gencyber_gencyber_2020} is a camp program funded by the NSF and NSA where universities host K-12 students and attempt to bring awareness of and excitement for cybersecurity. We note that our university also conducts GenCyber summer camps where, in previous camps, cryptographic algorithms and schemes were taught using the Python programming language. We will be switching to CryptoScratch in our future camps to simplify the cryptography concepts for students as Scratch is a platform that many K-12 students express as the most intuitive.

Other initiatives such as Hacker High school from the Institute for Security and Open Methodologies (ISECOM)~\cite{isecom_hacker_2021} provide a self-directed learning curriculum focused on attack and defense skills. The AFA CyberPatriot program created by the Air Force Association \cite{northrop_grumman_foundation_what_2013} is a program that works to inspire students to pursue careers in cybersecurity or related STEM fields. Other examples include \cite{ledeczi_netsblox_2019} which is a week-long summer camp program for high school students addressing a broad range of cybersecurity techniques. The private sector has also made efforts to improve cybersecurity education and awareness. For example, the Bits N Bytes organization \cite{bits_n_bytes_bits_2020} is dedicated to bringing awareness of the field of cybersecurity to young students. Cybrary is a platform that provides self-paced training in cybersecurity for a wide range of skill levels \cite{cybrary_free_2021}. Another example is \cite{paper-baciu} which presents an online learning tool for attack and defense. Works such as Cybersecurity Lab as a Service (CLaaS) \cite{tunc_claas_2015} have also looked at creating services that can run other cybersecurity tools on their platform. We note that even though these initiatives focus on diverse cybersecurity domains, they include cryptography as one of the core components. Our analysis of these initiatives reveals that most focus on simple tools for demonstrating simple cryptographic algorithms such as Caesar cipher. In the case of complex algorithms, the education is primarily verbal and does not provide tools for hands-on practice of cryptography. The CryptoScratch framework can fill the gap for these initiatives.

Some initiatives also focus on cryptography such as the Cryptographic Protocols sections of Classical CS Unplugged~\cite{computer_science_education_research_group_at_the_university_of_canterbur_cryptographic_nodate}. Similarly, NCC Group hosts the Cryptopals Crypto Challenges~\cite{ncc_groups_cryptography_services_cryptopals_nodate} that provide participants the opportunity to write code to solve cryptographic problems. Cryptopals focuses on already strong programmers who want to explore cryptographic concepts independently. The CryptoClub Project \cite{beissinger_about_2021} provides curriculum for teachers and an interactive website \cite{university_of_chicagos_center_for_stem_education_cryptoclub_2016} that allow the exploration of classical cryptographic techniques. LLCipher from the Lincoln Laboratory \cite{mit_lincoln_laboratory_llcipher_2021} is a workshop designed to teach the theory of Cryptography to students interested in the mathematics of cybersecurity. They focus on teaching the use of classical cryptography and the theory of cryptography. Prior works on teaching cryptography to middle-school students have also used various techniques. For instance, \cite{lindmeier_keeping_2020} used drawings to help students learn and explain the basics of cryptography and the components of secure communications systems. González-Tablas et al.~\cite{gonzales20} used a CryptoGo card game to teach cryptography. Other works such as \cite{hutchison_teaching_2010} attempted to develop visual techniques to teach public key infrastructure concepts. These works differ from ours as they either provide tools to teach the concepts or have specific use-cases that cannot be extended to learn other concepts. Meanwhile, CryptoScratch provides a generic, simple and visual tool for understanding and practicing several cryptographic algorithms and schemes. As such, CryptoScratch can be directly integrated into the above initiatives and curriculum. For instance, Cryptopals can use the CryptoScratch Task Block to attract K-12 students on their platform who may not already be strong programmers but are interested in solving complex cryptographic problems.

Several research papers also showcase the importance of cryptography education and tools for improving cybersecurity education and awareness. For instance, Svabensky~et~al.~\cite{10.1145/3328778.3366816} published a systematic literature review of SIGCSE and ITiCSE conference papers from 2010 to 2019. They reported that only a small portion of the papers focused on cryptography among all sub-fields of cybersecurity education across all age groups. In particular, works focused on K-12 students were limited. While there is a gap between ~\cite{10.1145/3328778.3366816} and now, cryptography in K-12 cybersecurity education is still underrepresented. There are also some excellent prior works to be noted, such as teaching cryptography using an interactive protocol called CYPHER~\cite{Younis2020}, research on secure coding with misuse pattern detection techniques~\cite{CryptoTutor} and secure coding for K-12 female students~\cite{Chattopadhyay2019}.  In \cite{da_silva_criptolab_2018}, the authors implemented a game where students used cryptography and cryptanalysis to solve problems and then used the clues to move to the next clue station. These works highlight the importance of tools for hands-on practice of cryptography that our CryptoScratch framework is intended to provide.

\section{CryptoScratch Design and Implementation}\label{sec:design}

This section focuses on the design and implementation details of CryptoScratch. In particular, we outline our reasons for implementing CryptoScratch on the Scratch platform, the design of modern cryptographic algorithms as CryptoScratch Blocks, the design of the Task Block for nudging learners towards more secure solutions, and details of the challenges implemented to work with the aforementioned Task Block.

\subsection{Developing CryptoScratch on Scratch}
The CryptoScratch framework is implemented as an extension to the Scratch programming language. We chose Scratch as the underlying environment for multiple reasons. First, Scratch is widely available via the Internet as an application-as-a-service or as a locally installed application for Windows, macOS, ChromeOS, and Android~\cite{scratch_foundation_scratch_nodate-1}. The environment is actively maintained by the Lifelong Kindergarten Group in the Media Laboratory at the Massachusetts Institute of Technology (MIT) through the Scratch Foundation \cite{scratch_foundation_scratch_nodate}. Due to this wide availability of Scratch to anyone with even occasional access to the Internet, we believe that the framework and the implemented challenges can be immediately available to groups with diverse socioeconomic backgrounds, including underrepresented groups.

Second, due to its popularity, the Scratch programming language is used in many K-12 schools in the US and the world~\cite{scratch_foundation_scratch_2020}. It is well-known for its visual block-based interface, intuitive environment, and features that make it easy for students to understand complex computer science constructs. We believe that the existing use of Scratch in the K-12 curriculum will help the adoption of CryptoScratch, as educators likely already have familiarity with Scratch making it easier for them to incorporate CryptoScratch into their teaching. Making the framework available to existing Scratch users will potentially increase its awareness and awareness of cryptographic algorithms, schemes, and protocols. As the platform is free for all, we note that it is also a financially sustainable solution to teach cybersecurity at the K-12 level in schools from underprivileged areas.

Third, Scratch currently has several extensions related to educational modules such as Artificial Intelligence, Machine Learning, and Embedded Programming. Currently, these efforts have not yet included cybersecurity even though several works \cite{jin_2018,info11020121,10.1145/3287324.3287450,10.1145/3328778.3366878,10.1145/3282844,baciu-ureche_adventures_2019,9368440} demonstrate that we can effectively teach cybersecurity concepts through gamification and visual feedback (both features of Scratch). This work intends to address this gap by implementing the first Scratch extension for cybersecurity dedicated to cryptography. In the future, we intend to further extend Scratch incorporating other vital aspects of cybersecurity, such as firewalls and intrusion detection systems. The incorporation of CryptoScratch on Scratch not only enables students to use modern cryptographic algorithms easily but also enables them to combine these algorithms to form complex cryptographic schemes like PGP (see Figure \ref{fig:CryptoScratchPGPCode}).

\subsection{Designing blocks for cryptographic algorithms}

\begin{figure}[t]
  \centering
  \includegraphics[width=\columnwidth]{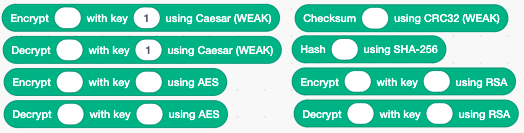}
  \caption{CryptoScratch blocks for the implemented algorithms}
  \label{fig:AlgorithmsList}
\vspace{-.5em}
\end{figure}

One of the primary goals of CryptoScratch is to introduce students to modern and secure cryptographic algorithms using a visual and intuitive interface and without focusing on the mathematics behind the algorithms. The intention behind this goal is that, at the middle-school level, students should not be expected to understand the inner workings of algorithms such as AES or RSA. Instead, they should be made aware of different types of cryptographic algorithms, the security services they provide, and use-cases. The design of the visual blocks allows for a simplified discussion by focusing on the application of cryptography while avoiding the mathematical premises of the algorithms.

\begin{figure}[t]
  \centering
  \includegraphics[width=0.8\columnwidth]{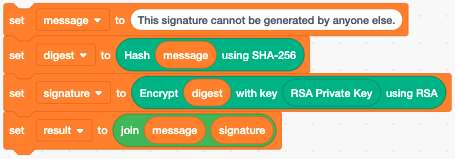}
  \caption{Example Scratch code demonstrating the creation of Digital Signature using CryptoScratch}
  \label{fig:Signature}
\vspace{-1em}
\end{figure}

The current implementation of CryptoScratch supports the following cryptographic algorithms - Caesar, AES, CRC32, SHA-256, and RSA. \Cref{fig:AlgorithmsList} shows the default setup of the implemented blocks. We note that this set comprises both weaker and stronger cryptographic algorithms to enable students to understand the inherent security problems with weak algorithms and how the more secure modern algorithms provide a higher level of security. The industry-standard AES and RSA encryption algorithms demonstrate symmetric and asymmetric cryptography, respectively. We note that the RSA algorithm also implements blocks that enable students to generate an RSA public and private key pair. The Caesar cipher can explain cryptanalysis attacks on symmetric cryptosystems. In addition, CRC32 and SHA-256 can illustrate the concepts of one-way functions, error checking, and digital signatures, with CRC32 also providing the capability of showing collisions. We note that the development of CryptoScratch is an ongoing effort, and many other algorithms/schemes (e.g., DES, 3DES, HMAC, Diffie Hellman, ECC) will be added in the future versions of the framework.  We emphasize that the framework is still practical as students can develop complex cryptographic schemes by combining the existing blocks in different ways. An example of such a  scheme is Digital Signatures that combines the SHA-256 and RSA blocks as shown in \Cref{fig:Signature}.

Within Scratch, we have implemented all the cryptographic blocks as JavaScript functions (ES6) with support for most modern browsers. Internally, these functions invoke well-known cryptographic implementations provided by the \emph{crypto-browserify} npm package~\cite{crypto_browserify}. We note that the CryptoScratch extension is fully compatible with Scratch since it uses the guidelines provided by the Scratch community for creating new extensions \cite{scratch_extensions}, so that no changes are required to the Scratch platform and source code. 

\lstset{
   language=JavaScript,
   extendedchars=true,
   basicstyle=\scriptsize\ttfamily,
   showstringspaces=false,
   showspaces=false,
   numbers=left,
   numberstyle=\scriptsize,
   numbersep=8pt,
   tabsize=2,
   breaklines=true,
   showtabs=false,
   captionpos=b
}

\begin{center}
\begin{tabular}{c}
\begin{lstlisting}[frame=single,basicstyle=\scriptsize,caption={Example code snippet of AES Encryption Block},captionpos=b,language=JavaScript,label={lst:aes},linewidth=8.6cm]
const AES = require("browserify-aes");
...
   opcode: "doAesEncrypt",
   blockType: BlockType.REPORTER,
   text: "Encrypt [PLAINTEXT] with key [KEY] using AES",
   arguments: {
      ...
   }
...

doAesEncrypt (args) {
   const plaintext = args.PLAINTEXT;
   const key = convertToKey(args.KEY);
   
   var aes = AES.createCipheriv("aes-128-ecb", key, null);
   aes.setAutoPadding(true);

   var ciphertext = aes.update(plaintext, "utf8", "hex");
   ciphertext += aes.final("hex");
   
   return ciphertext;
}
...
\end{lstlisting}
\end{tabular}
\end{center}

Listing \ref{lst:aes} shows code snippets of the implementation of the `\texttt{Encrypt [PLAINTEXT] with key [KEY] using AES}’ block (shown in \Cref{fig:AlgorithmsList}, row 3). Note that we have removed other code here to focus specifically on the AES Encryption process. In this example, the text `\texttt{Encrypt \_ with key \_ using AES}’ denotes the text displayed in the block. Observe that the \texttt{[PLAINTEXT]} and \texttt{[KEY]} parameters are replaced by user input fields. Internally, these parameters render as string arguments to the JavaScript function defined by the \emph{opcode} parameter. The \emph{opcode} parameter, in our example, points to the \texttt{doAesEncrypt} JavaScript function that gets executed when the specific block gets executed. We also note that we have implemented all the CryptoScratch blocks as \texttt{REPORTER} blocks as all the algorithms accept some inputs and return a value. More precisely, the example block internally gets translated to \texttt{doAesEncrypt([PLAINTEXT], [KEY])} which returns the ciphertext in hex format. The \texttt{doAesEncrypt} JavaScript function then invokes the appropriate \emph{browserify} objects and functions on the arguments to perform AES encryption in ECB mode to produce the ciphertext. The given key/password is hashed to create a 128-bit key to be used with the AES algorithm. We chose to use a 128-bit key size to reduce computing power on the Scratch platform. The other blocks for AES decryption, RSA encryption and decryption, SHA-256 hashing, Caesar Cipher, and CRC32 were also implemented using the same methodology described for AES Encryption and are not discussed further for the sake of brevity.

\subsection{Designing a Task Block for automated code feedback}\label{sec:task_block}

\begin{figure}[t]
  \centering
  \includegraphics[width=.5\linewidth]{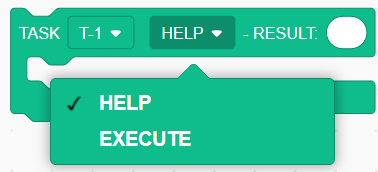}
  \caption{CryptoScratch Task Block to enable users to get instructions and test the results of tasks from within CryptoScratch}
  \label{fig:task_block}
\vspace{-1em}
\end{figure}

To the best of our knowledge, the Scratch platform does not define any blocks or functions to provide feedback to users on their code implementation. We believe that giving feedback is especially important in cryptography since mistakes in cryptographic implementations can often lead to critical security vulnerabilities. To address this gap, we have designed and implemented a new Task Block in CryptoScratch that serves two purposes. First, this block can be used as a generic wrapper to present various challenges that learners can solve to learn cryptography. We refer the readers to Section \ref{sec:tasks} for a list of challenges already implemented using the Task Block. Second, using the block to encapsulate the learner’s solution for the challenge enables this block to check insecure implementations of cryptographic algorithms and the student's final solution. This task block is shown in Figure \ref{fig:task_block} where users can select a task that they'd like to solve, and then choose to get help or execute the task\footnote{A video of the task block and code checker is available at \url{ https://www.dropbox.com/s/235w7posvwh8bl1/CryptoScratch.mov}}.

\begin{figure}[t]
    \centering
    \begin{subfigure}[b]{0.4\columnwidth}
        \centering
        \includegraphics[width=\columnwidth]{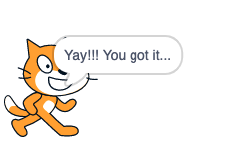}
        \caption{Task success message}
        \label{fig:SpriteSuccess}
    \end{subfigure}%
    ~ 
    \begin{subfigure}[b]{0.4\columnwidth}
        \centering
        \includegraphics[width=\columnwidth]{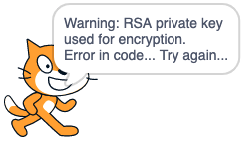}
        \caption{Incorrect usage warning}
        \label{fig:SpriteWarning}
    \end{subfigure}
    \caption{Examples of feedback provided by the CryptoScratch Task Block}
\vspace{-1em}
\end{figure}

The intuition behind the feedback capability of the Task Block can be demonstrated using the popular Pretty Good Privacy (PGP) scheme. Before discussing the scheme, we will introduce some notations to simplify our discussion. Let $K\{M\}$ denote the process of encrypting a secret message $M$ using a symmetric key $K$ shared between two users A and B. In the case of public-key cryptography, let $\{M\}_B$ denote that the secret message $M$ is encrypted using user B’s public key and $[M]_A$ denote that message $M$ is encrypted using user A’s private key. Using these notations, the PGP scheme for protecting the confidentiality of a large message $M$ is represented as $K\{M\}|\{K\}_B$. Here, sender A generates a random symmetric key $K$ to encrypt the message $M$. To send the key $K$ to receiver B, sender A encrypts $K$ using B’s public key. This scheme provides confidentiality given that an attacker cannot decrypt the key $K$ and the message $M$ without knowledge of B’s private key. If a student implements the correct solution, the Task Block checks the accuracy of their solution by (1) decrypting $\{K\}_B$ first to obtain $K$ using B's private key, (2) decrypting $M^{’}=K\{M\}$ to obtain the message $M^{’}$, and (3) verifying that the challenge message $M$ and the decrypted message $M^{’}$ are the same. For a correct solution, the Sprite displays a success message to the student as shown in Figure \ref{fig:SpriteSuccess}. However, a common mistake for learners new to cryptography stems from confusion between using the private/public key for encryption. If a learner makes a mistake and uses the wrong key for encryption, the scheme will become $K\{M\}|[K]_A$. This scheme no longer provides confidentiality as an attacker can use A’s public key to decrypt the key $K$ and then use $K$ to decrypt the secret message $M$. The Task Block is designed to display a warning to the learner as shown in Figure \ref{fig:SpriteWarning}, to make them aware of the mistake and nudge them towards a more secure solution.   

\begin{figure}[t]
    \centering
    \includegraphics[width=0.4\columnwidth]{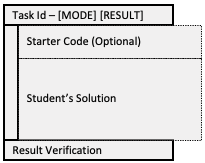}
    \caption{\label{fig:TaskBlock}Task Block setup structure}
    \vspace{-1em}
\end{figure}

The PGP example above showcases one security problem (i.e., confidentiality breach) that the Task Block can detect. We note that the design of this block enables the detection of many different forms of security problems. Let’s take another example of a Digital Signature of a message $M$ that can be represented as $M|[H(M)]_A$, where $H(M)$ is the hash of message $M$ using a secure hash function $H$. In this specific case, the Task Block can detect problems such as a learner using a public key instead of a private key for signing the message $M|\{H(M)\}_B$ (authentication flaw); as well as cases where a learner may use a collision-prone insecure function such as CRC32 (signature spoofing). We note that the above checks and  others are already implemented into CryptoScratch, and we intend to keep adding more checks in future revisions. 

To make the above checks feasible, the Task Block has specific sections as shown in \Cref{fig:TaskBlock}. The top section of the block `\texttt{Task Id - [MODE] [RESULT]}' initializes the task by setting up the parameters required for the successful completion of a task. Currently, the \texttt{[MODE]} parameter supports two execution modes - `HELP' to display help information to the student for solving the task and `EXECUTE' that executes the student's solution for the task. The \texttt{[RESULT]} parameter is a placeholder where students can place the variable that will store the final solution. The bottom section of the block `\texttt{Result Verification}' performs the code checking for secure usage and validates the \texttt{[RESULT]} variable by matching the task's solution with the learner’s solution. For instance, in the case of the PGP scheme in Figure \ref{fig:CryptoScratchPGPCode}, the code checker performs decryption of the student's solution by using the PGP decryption process described before and checks whether the decrypted text matches the text specified in the task. Everything between the top and bottom sections is the learner’s solution to solve the challenge identified by the task. We note that we will also provide optional starter codes that students will download from the project’s public website.



\subsection{Designing cryptographic challenges on CryptoScratch}\label{sec:tasks}
The CryptoScratch framework currently implements eight challenges designed to work with the Task Block so that students can get appropriate feedback as they solve the challenges. We plan to make the CryptoScratch source code and all the challenges publicly available on GitHub to encourage further contributions to this challenge set. All these challenges are designed to help students learn about cryptography. They present situations similar to real-world systems, albeit simplified, where cryptography offers a viable solution to secure sensitive information.

\begin{figure}[t]
  \centering
  \includegraphics[width=0.8\columnwidth]{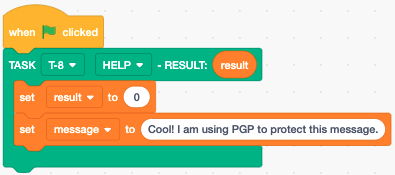}
  \caption{Example Task Block with Starter code for a Pretty Good Privacy implementation task}
  \label{fig:starter_code}
  \vspace{-1em}
\end{figure}

The current set of challenges range from encrypting sensitive data using AES \texttt{(Task 1)} and decrypting sensitive data using AES \texttt{(Task 2)} to demonstrate the power of symmetric cryptography, known plaintext-ciphertext cryptanalysis attack on the Caesar cipher \texttt{(Task 3)} to demonstrate problems with weak cryptographic algorithms, encrypting sensitive data using RSA \texttt{(Task 4)}, and decrypting sensitive data using RSA \texttt{(Task 5)} to demonstrate the power of asymmetric cryptography, and hashing data using the SHA-256 algorithm \texttt{(Task 6)} to demonstrate the concept of message integrity. As the above challenges focus on single cryptographic algorithms, we have created two additional challenges to demonstrate to learners how to combine cryptographic algorithms to create complex cryptographic schemes. These challenges are creating and appending a digital signature to a message \texttt{(Task 7)} and protecting the confidentiality of a message using Pretty Good Privacy (PGP) \texttt{(Task 8)}, both discussed in~\Cref{sec:task_block}. We note that, for the K-12 level, Tasks 7 and 8 provide sufficient challenge as (1) learners must understand the differences between symmetric cryptography, asymmetric cryptography and secure hashing algorithms, (2) learners must understand the purpose behind these cryptographic schemes, and then (3) learners must implement them in CryptoScratch by combining together several cryptographic blocks (AES, RSA, SHA-256). We have also created starter codes for each challenge that students can download from our public website for CryptoScratch and upload on the Scratch environment. Figure \ref{fig:starter_code} shows an example starter code for the PGP task\footnote{The screenshots of all tasks starter code and their solutions is available at \url{https://www.dropbox.com/sh/p7xmtpgkwaug77f/AADaLKjl9h2P8H2Pz84Qy1Baa}} whose solution is shown in Figure \ref{fig:CryptoScratchPGPCode}. As we add more blocks to CryptoScratch, we will add more challenges (e.g., Message Authentication Codes, Authenticated Encryption) to enable students to learn the basics of these algorithms and then combine them to create additional functionality using the visual interface of CryptoScratch.


\section{CryptoScratch Evaluation}\label{sec:userstudy}

This section outlines the details of the user study with 16 middle-school students and the qualitative and quantitative results obtained from the study.

\subsection{Design of the User Study}
To better understand the impacts of CryptoScratch on K-12 students’ understanding of cryptography and its application to cybersecurity, we conducted a 3-day workshop with middle-school students\footnote{This research was approved by the authors' university's Institutional Review Board (IRB).}. The details of the workshop, the participants and the study are given below.

\vspace{.2em}
\noindent \textbf{\textit{Recruitment Method:}} We utilized a snowball sampling method starting with the immediate contacts of the research team to connect with parents of middle-school students. Using this method, 16 middle school students registered and attended the workshop. These students completed a pre-workshop survey, a post-workshop survey and submitted their solutions for the challenges as requested. We refer the readers to~\Cref{sec:tasks} for the list of challenges given to these students.

\vspace{.2em}
\noindent \textbf{\textit{Students Distribution:}} Table \ref{tab:sample_grade} shows the grade and gender distribution of the students. The workshop students ranged from sixth grade to tenth grade, of which the 8th and 9th grades accounted for 50\% of all the students. Among the students, 56.25\% were boys and 43.75\% were girls.

\begin{table}[h]
    \small
    \caption{Students' Grade and Gender Distribution}
    \label{tab:sample_grade}
    \centering
    \begin{tabular}{c|c|c}
        \toprule
        \textbf{Grade} & \textbf{N} & \textbf{\%}\\
        \midrule
        6 &	3 & 18.75\\
        7 & 4 & 25.00\\
        8 & 5 & 31.25\\
        9 & 3 & 18.75\\
        10 & 1 & 6.25\\
        \bottomrule
    \end{tabular}\hspace{3em}
    \begin{tabular}{c|c|c}
        \toprule
        \textbf{Gender} &  \textbf{N} & \textbf{\%}\\
        \midrule
        Boy & 9 & 56.25\\
        Girl & 7 & 43.75\\
        \bottomrule
    \end{tabular}
\end{table}

\vspace{.2em}
\noindent \textbf{\textit{Training / Workshop:}} The recruited middle school students were given a 9-hour (3 hours * 3 days) hands-on training session (virtual) on cryptography and the CryptoScratch framework. Each day, students were given a combination of instructions on the basics of cryptography such as symmetric cryptography, asymmetric cryptography, hashing, and hands-on challenges to complete using CryptoScratch. Multiple instructors were available while students were working on the challenges to answer any general questions and solve any technical problems. We note that the first six challenges helped the students better understand how the recently introduced cryptographic concepts worked. The next two (PGP and Digital Signature) were used to assess the students' understanding of the cryptographic concepts covered and their ability to utilize CryptoScratch to implement complex solutions requiring the combination of techniques that were taught.

\subsection{Pre- and Post-Survey Evaluation of CryptoScratch}

Our goal with the pre- and post-survey was to understand the workshop’s impact on the students’ inclination towards cryptography. In particular, we wanted to assess whether students enjoyed learning about cryptography and saw solving challenges using CryptoScratch as a fun activity. Using the results, we wanted to determine if it is possible to generate curiosity among K-12 students towards cybersecurity so that they consider cybersecurity as a future profession.

\vspace{.2em}
\noindent \textbf{\textit{Pre- and Post-surveys:}} Before starting the workshop, we obtained documented parental consent.  We then asked the students to complete a research participation assent form and pre-training survey to share their demographic information, programming comfort level, experience with Scratch, and experience with cryptography. During the workshop, we also asked them to submit solutions for specific challenges solved using CryptoScratch.

After the workshop completed, the students completed a post-training survey to provide feedback on CryptoScratch. For the post-survey, the students were asked to evaluate if CryptoScratch helped in their learning, whether they noticed improvement in their understanding of cryptography, how they found the challenges in terms of clarity and difficulty, and whether the challenges were fun to solve using CryptoScratch.

\vspace{.2em}
\noindent \textbf{\textit{Analysis of Survey Results:}} The pre-survey results indicated that most of the students were new to programming and had limited experience with Scratch. For instance, \Cref{tab:scratch} shows that only 5 (31.25\%) students had some experience with Scratch, with only one student having significant experience with the platform. We note that six students started the workshop with no programming and Scratch experience and started learning both Scratch and CryptoScratch only during the workshop. Nearly all the 16 students had low to no experience with cryptographic algorithms and schemes.

\begin{table}[h]
    \small
    \caption{Scratch Experience}
    \centering
    \label{tab:scratch}
    \begin{tabular}{l|c|c}
        \toprule
        Scratch Usage & N & \%\\
        \midrule
        Never & 6 & 37.50\\
        A little & 5 & 31.25\\
        Some & 4 & 25.00\\
        A lot & 1 & 6.25\\
        \bottomrule
    \end{tabular}
    \vspace{-1em}
  \end{table}

The post-survey results indicated a positive experience of cryptography among many students. For example, 15 (93.75\%) students out of the 16 reported agreement with the statement that they gained more knowledge about cryptography due to the workshop. Meanwhile, one student expressed that they were neutral about the above statement. The results also indicated a positive experience with the CryptoScratch framework and Task Block. For instance, 11 (68.75\%) students reported that they found the framework easy to use and understand, and 3 (18.75\%) students said they were neutral about the above statement. In contrast, 2 (12.5\%) students reported not finding CryptoScratch easy to use and understand. We remind the readers that, before the workshop, 11 students mentioned that they had little to no experience with Scratch. This means that at least six of these students found the CryptoScratch framework easy to use and understand despite their limited experience with the Scratch platform.





\Cref{tab:labs} shows the results related to students’ understanding of the challenges. We note that question Q1 checks whether students found the challenges easy to complete, Q2 checks whether students found the challenges easy to understand, and Q3 checks whether students had fun solving the challenges. We observe that the responses are mainly positive, indicating that the students had a positive experience with the challenges. For instance, 9 (56.25\%) students found the challenges to be easy to complete, 5 (31.25\%) were neutral, and only 2 (12.5\%) students found the challenges not as easy. We remind the readers that these challenges comprise two (Pretty Good Privacy and Digital Signatures) that require a deeper understanding of cryptographic algorithms and schemes and are considered hard to comprehend for students even at the university level. For Q2, a majority (13 students) reported that the challenges were easy to understand, which reveals a firmer understanding of cryptographic concepts. Meanwhile, 14 students reported that they had fun doing the challenges, while two were neutral about that statement. We note that these results indicate that students find CryptoScratch easy to understand and fun to work with and reaffirm that the tool can be effective to teach cryptography in K-12 classrooms.

\begin{table}[h]
    \small
    \caption{Experience with CryptoScratch Challenges}

    \centering
    \label{tab:labs}
    \begin{tabular}{l|c|c|c}
        \toprule
        Student Response & Q1 & Q2 & Q3 \\
        \midrule
        Completely disagree	&   0	&   0   &  0\\
        Somewhat disagree   &   2	&   1   &  0\\
        Neutral	            &   5   &	2   &  2\\
        Somewhat agree	    &   4   &	7   &  7\\
        Completely agree	&   5   &	6   &  7\\
        \bottomrule
    \end{tabular}
\end{table}



We note that the sample size of this study is not sufficient to produce any statistically significant results. However, we also emphasize that these initial results encourage additional studies to evaluate the effectiveness of CryptoScratch further. A larger study of more diverse student groups (e.g., sociodemographic groups, of technology access and comfort levels, and previous programming experience) will be performed in the future with enough participants to allow a generalizable, statistically significant conclusion.

\subsection{Evaluation of Student Solutions}

This section performs a quantitative analysis of CryptoScratch by analyzing the solutions to the challenges that students completed during the workshop. We note that, among the eight challenges (\Cref{sec:tasks}) completed by the students, our discussion focuses on two challenges - Pretty Good Privacy and Digital Signatures. This is because the first six challenges focused on simple usages of cryptographic algorithms such as AES, RSA, Caesar, and SHA-256 that nearly all the students could complete quickly. The PGP and Digital Signature challenges, on the other hand, were deemed harder by the students, and not all the students were able to complete them in the allocated time of 20 minutes. As such, these challenges provide us insights into whether the students gathered a solid foundation of cryptography and were able to use them to solve cryptographic problems in CryptoScratch.


\begin{figure}[t]
  \centering
  \includegraphics[width=0.75\columnwidth]{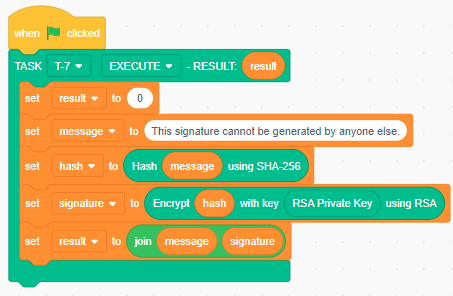}
  \caption{Example of a correctly implemented Digital Signature solution by a Student}
  \label{fig:digital_sig_student_success}
  \vspace{-1em}
\end{figure}

In the case of the PGP and Digital Signature challenges, we note that we provided the students with starter code such as the one shown in~\Cref{fig:starter_code}. 
Our deadline to complete each challenge was 20 minutes, an assessment by the instructors based on the completion time of the easier challenges.
For Task 7 (Digital Signature), an analysis of the submissions found that 8 (66.67\%) out of 12 students implemented a correct solution similar to \Cref{fig:digital_sig_student_success}. Two (16.67\%) students submitted an incorrect solution, and 2 (16.67\%) students submitted the starter code given to them. Analyzing the two incorrect solutions revealed that the students were close to obtaining the correct results had we given them more time. For instance, one student’s error was to overwrite the original message with a hash of the message and then sign the hash of the hash. The other student submitted a partial solution where they successfully calculated the hash and signed the hash. The only additional step they needed to reach the correct solution was to concatenate the message with the signature in the result.


\begin{figure}[t]
  \centering
  \includegraphics[width=0.7\columnwidth]{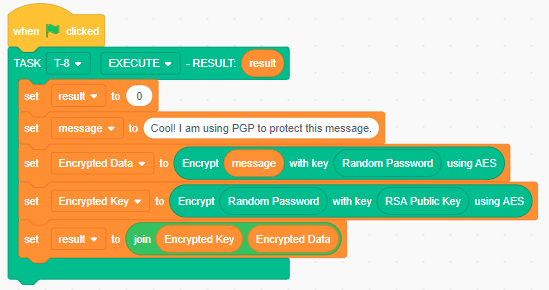}
  \caption{Example code of PGP with an error by the student}
  \label{fig:pgp_error}
  \vspace{-1.5em}
\end{figure}

\begin{figure}[t]
  \centering
  \includegraphics[width=0.7\columnwidth]{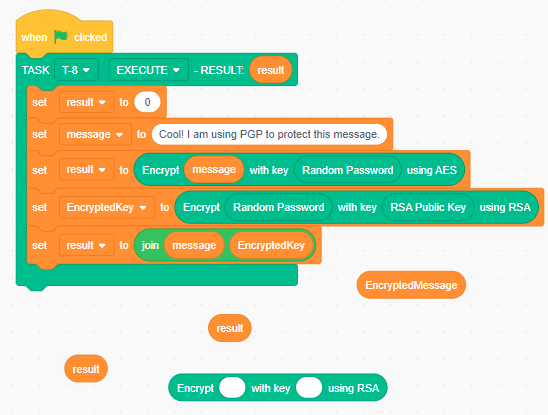}
  \caption{Example code of an incomplete PGP solution by a student}
  \label{fig:pgp_incomplete}
  \vspace{-1.5em}
\end{figure}

For Task 8 (PGP), 9 (60\%) out of 15 students submitted the correct solution, 4 (26.7\%) made some basic errors pointing to potential confusion between the uses of AES and RSA, and 2 (13.3\%) submitted the starter code given to them. \Cref{fig:pgp_error} shows the error made by two students. In their case, they tried to encrypt the random key with the RSA public key correctly but used the AES block instead of the RSA block for encryption. We note that this error could also mean negligence from the student and may not specifically highlight confusion between symmetric and asymmetric cryptography. One solution (\Cref{fig:pgp_incomplete}), an ideal case for the Task Block, mistakenly sent the plaintext message instead of the ciphertext message. Again, this mistake could be an artifact of the time constraints on the student rather than confusion between plaintext and ciphertext. Our intuition is derived from the fact that the student did think about encrypting the message (as suggested by the EncryptedMessage variable block). Our planned larger study with more students will look at these errors more closely and understand the underlying causes of the errors. We plan to use the analysis further to improve the feedback capability of the Task Block.

\section{Conclusion \& Future Work}\label{sec:conclusion}

This paper presented the design, implementation, and evaluation of a new framework,  CryptoScratch, built as an extension of the Scratch programming language. The CryptoScratch framework implements strong cryptographic algorithms (e.g., AES, RSA, SHA-256) and weaker algorithms (e.g., Caesar cipher, CRC32) as simple, visual, and intuitive blocks. These blocks enable students to interact with cryptographic algorithms, learn weaknesses and strengths of the algorithms, and use the blocks to create more complex cryptographic schemes such as Pretty Good Privacy (PGP) and Digital Signatures. The successful development of CryptoScratch demonstrates that a visual block-based tool for teaching cryptography without the mathematical complexity is possible (RQ1).  By using the CryptoScratch environment for an initial study, the ability to use CryptoScratch to help K-12 students understand and complete basic implementations of modern cryptographic systems has been demonstrated (RQ2).

We believe that the framework enables hands-on practices for cryptography and can be directly integrated into existing initiatives (e.g., GenCyber, CryptoClub) and the K-12 curriculum. We also presented the design and implementation of a new Task Block on CryptoScratch that enables students to solve cryptographic challenges to learn about cryptography. The Task Block generates feedback for student solutions to nudge them towards more secure cryptographic solutions. We note that the development of CryptoScratch is an ongoing effort, and other algorithms/schemes (e.g., HMAC, Diffie Hellman, ECC) will be added to the framework. We plan to make CryptoScratch and the Task Block open-source and publicly available through the MIT Scratch website. 



We also evaluated CryptoScratch through an initial usability study with 16 middle-school students. During the study, we introduced the students to cryptography and the CryptoScratch framework. We asked them to complete a set of challenges to learn about cryptographic algorithms and then use those algorithms to implement complex cryptographic schemes. The study indicated that nearly 60\% of the students could quickly grasp and implement complex cryptography concepts using CryptoScratch, while nearly 90\% showed comfort with cryptography concepts and use-cases. We plan to do a larger study of more diverse student groups and enough participants to allow a generalizable, statistically significant conclusion.

\bibliographystyle{IEEEtran}
\bibliography{paper}

\begin{thebibliography}{10}
\providecommand{\url}[1]{#1}
\csname url@samestyle\endcsname
\providecommand{\newblock}{\relax}
\providecommand{\bibinfo}[2]{#2}
\providecommand{\BIBentrySTDinterwordspacing}{\spaceskip=0pt\relax}
\providecommand{\BIBentryALTinterwordstretchfactor}{4}
\providecommand{\BIBentryALTinterwordspacing}{\spaceskip=\fontdimen2\font plus
\BIBentryALTinterwordstretchfactor\fontdimen3\font minus
  \fontdimen4\font\relax}
\providecommand{\BIBforeignlanguage}[2]{{%
\expandafter\ifx\csname l@#1\endcsname\relax
\typeout{** WARNING: IEEEtran.bst: No hyphenation pattern has been}%
\typeout{** loaded for the language `#1'. Using the pattern for}%
\typeout{** the default language instead.}%
\else
\language=\csname l@#1\endcsname
\fi
#2}}
\providecommand{\BIBdecl}{\relax}
\BIBdecl

\bibitem{Isaca_ransomware}
\BIBentryALTinterwordspacing
ISACA, ``{Surge} in {Ransomware} and 10 {Biggest} {Attacks} in {2021},'' 2021.
  [Online]. Available:
  \url{https://www.isaca.org/resources/news-and-trends/newsletters/atisaca/2021/volume-35/surge-in-ransomware-attack-and-10-biggest-attacks-in-2021}
\BIBentrySTDinterwordspacing

\bibitem{isc2_cybersecurity_2020}
\BIBentryALTinterwordspacing
(ISC)\textsuperscript{2}, ``Cybersecurity {Professionals} {Stand} {Up} to a
  {Pandemic}: ({ISC})\textsuperscript{2} {Cybersecurity} {Workforce} {Study},
  2020,'' Tech. Rep., 2020. [Online]. Available:
  \url{https://www.isc2.org/-/media/ISC2/Research/2020/Workforce-Study/ISC2ResearchDrivenWhitepaperFINAL.ashx}
\BIBentrySTDinterwordspacing

\bibitem{isaca_state_2021}
\BIBentryALTinterwordspacing
ISACA and H.~Technologies, ``State of {Cybersecurity} 2021, {Part} 1: {Global}
  {Update} on {Workforce} {Efforts}, {Resources} and {Budgets},'' Tech. Rep.,
  2021. [Online]. Available:
  \url{https://www.isaca.org/go/state-of-cybersecurity-2021}
\BIBentrySTDinterwordspacing

\bibitem{gencyber_gencyber_2020}
\BIBentryALTinterwordspacing
GenCyber, ``{GenCyber},'' 2021. [Online]. Available:
  \url{https://www.gen-cyber.com/}
\BIBentrySTDinterwordspacing

\bibitem{northrop_grumman_foundation_air_nodate}
\BIBentryALTinterwordspacing
{Northrop Grumman Foundation}, ``Air {Force} {Associations} {CyberPatriot}.''
  [Online]. Available: \url{https://www.uscyberpatriot.org/}
\BIBentrySTDinterwordspacing

\bibitem{nist_national_2016}
\BIBentryALTinterwordspacing
NIST, ``National {Initiative} for {Cybersecurity} {Education} ({NICE}),'' 2016.
  [Online]. Available:
  \url{https://www.nist.gov/itl/applied-cybersecurity/nice}
\BIBentrySTDinterwordspacing

\bibitem{pltw_pltw_2021}
\BIBentryALTinterwordspacing
PLTW, ``{PLTW} {Computer} {Science} {Curriculum} - {Cybersecurity},'' 2021.
  [Online]. Available:
  \url{https://www.pltw.org/our-programs/pltw-computer-science-curriculum}
\BIBentrySTDinterwordspacing

\bibitem{cyberorg_state_2020}
\BIBentryALTinterwordspacing
cyber.org, ``The {State} of {Cybersecurity} {Education} in {K}-12 {Schools},''
  2020. [Online]. Available:
  \url{https://cyber.org/sites/default/files/2020-06/The\%20State\%20of\%20Cybersecurity\%20Education\%20in\%20K-12\%20Schools.pdf}
\BIBentrySTDinterwordspacing

\bibitem{joint_task_force_on_cybersecurity_education_cybersecurity_2017}
\BIBentryALTinterwordspacing
{Joint Task Force on Cybersecurity Education}, ``Cybersecurity {Curricula}
  2017,'' 2017. [Online]. Available:
  \url{https://www.acm.org/binaries/content/assets/education/curricula-recommendations/csec2017.pdf}
\BIBentrySTDinterwordspacing

\bibitem{sabin_acmieee-cs_2016}
\BIBentryALTinterwordspacing
M.~Sabin, S.~Peltsverger, C.~Tang, and B.~M. Lunt, ``{ACM}/{IEEE}-{CS}
  {Information} {Technology} {Curriculum} 2017: {A} {Status} {Update},'' in
  \emph{Proceedings of the 17th {Annual} {Conference} on {Information}
  {Technology} {Education}}.\hskip 1em plus 0.5em minus 0.4em\relax ACM, 2016.
  [Online]. Available: \url{https://dl.acm.org/doi/10.1145/2978192.2978241}
\BIBentrySTDinterwordspacing

\bibitem{rowe_role_2011}
\BIBentryALTinterwordspacing
D.~C. Rowe, B.~M. Lunt, and J.~J. Ekstrom, ``The role of cyber-security in
  information technology education,'' in \emph{Proceedings of the 2011
  conference on {Information} technology education - {SIGITE} '11}, 2011.
  [Online]. Available:
  \url{http://dl.acm.org/citation.cfm?doid=2047594.2047628}
\BIBentrySTDinterwordspacing

\bibitem{kurt_peker_developing_2020}
Y.~Kurt~Peker and H.~Fleenor, ``Developing and {Implementing} a {Cybersecurity}
  {Course} for {Middle} {School},'' in \emph{National {Cyber} {Summit} ({NCS})
  {Research} {Track}}, K.-K.~R. Choo, T.~H. Morris, and G.~L. Peterson,
  Eds.\hskip 1em plus 0.5em minus 0.4em\relax Springer International
  Publishing, 2020.

\bibitem{university_of_chicagos_center_for_stem_education_cryptoclub_2016}
\BIBentryALTinterwordspacing
{University of Chicago’s Center for STEM Education}, ``{CryptoClub},'' 2016.
  [Online]. Available: \url{https://www.cryptoclub.org/}
\BIBentrySTDinterwordspacing

\bibitem{nsa_national_2020}
\BIBentryALTinterwordspacing
NSA, ``National {Cryptographic} {Museum}: {Virtual} {Programs},'' 2020.
  [Online]. Available:
  \url{https://www.nsa.gov/about/cryptologic-heritage/museum/tours#Virtual%20Programs}
\BIBentrySTDinterwordspacing

\bibitem{nsa_request_2021}
\BIBentryALTinterwordspacing
------, ``Request {NSA} activities, partners, and {STEM} fair judges for your
  {K}-12 classroom,'' 2021. [Online]. Available:
  \url{https://www.nsa.gov/resources/students-educators/k12-partnership/}
\BIBentrySTDinterwordspacing

\bibitem{scratch_foundation_scratch_nodate}
\BIBentryALTinterwordspacing
{Scratch Foundation}, ``Scratch - {Imagine}, {Program}, {Share}.'' [Online].
  Available: \url{https://scratch.mit.edu/}
\BIBentrySTDinterwordspacing

\bibitem{codebreakers_2020}
\BIBentryALTinterwordspacing
{Boston University Learning Resource Network}, ``{CodeBreakers} -
  {CyberSecurity} for {Young} {Women} in high school,'' 2020. [Online].
  Available: \url{https://www.bu.edu/lernet/cyber/}
\BIBentrySTDinterwordspacing

\bibitem{girlsgocyberstartwebsite}
\BIBentryALTinterwordspacing
{CyberStart America}, ``{GirlsGoCyberStart},'' 2021. [Online]. Available:
  \url{https://girlsgocyberstart.org/}
\BIBentrySTDinterwordspacing

\bibitem{womenscyberjutsu_2020}
\BIBentryALTinterwordspacing
{Women’s Society of Cyberjutsu}, ``{WSC} - {Women}{s}' {Society} of
  {Cyber}{jutusu},'' 2021. [Online]. Available:
  \url{https://womenscyberjutsu.org/page/WhoAreWe}
\BIBentrySTDinterwordspacing

\bibitem{NCL_2020}
\BIBentryALTinterwordspacing
{National Cyber League}, ``National {Cyber} {League} - {CyberSecurity}
  {Competition},'' 2021. [Online]. Available:
  \url{https://nationalcyberleague.org/home}
\BIBentrySTDinterwordspacing

\bibitem{isecom_hacker_2021}
\BIBentryALTinterwordspacing
ISECOM, ``Hacker {Highschool} : security awareness for teens,'' 2021. [Online].
  Available: \url{https://www.hackerhighschool.org/}
\BIBentrySTDinterwordspacing

\bibitem{northrop_grumman_foundation_what_2013}
\BIBentryALTinterwordspacing
{Northrop Grumman Foundation}, ``What is {CyberPatriot}?'' 2013. [Online].
  Available:
  \url{https://www.uscyberpatriot.org/Pages/About/What-is-CyberPatriot.aspx}
\BIBentrySTDinterwordspacing

\bibitem{ledeczi_netsblox_2019}
\BIBentryALTinterwordspacing
{\'A}.~Lédeczi, H.~Zare, and G.~Stein, ``Cybersecurity, women and
  minorities,'' in \emph{Proceedings of the 50th {ACM} {Technical} {Symposium}
  on {Computer} {Science} {Education}}.\hskip 1em plus 0.5em minus 0.4em\relax
  ACM, 2019. [Online]. Available:
  \url{https://dl.acm.org/doi/10.1145/3287324.3293749}
\BIBentrySTDinterwordspacing

\bibitem{bits_n_bytes_bits_2020}
\BIBentryALTinterwordspacing
{Bits N' Bytes}, ``Bits {N}' {Bytes} : {Our} {Mission},'' 2020. [Online].
  Available: \url{https://www.bitsnbytes.us.com/mission/}
\BIBentrySTDinterwordspacing

\bibitem{cybrary_free_2021}
\BIBentryALTinterwordspacing
Cybrary, ``Free {Cybersecurity} {Training} and {Career} {Development}
  ({Cybrary} {Homepage}),'' 2021. [Online]. Available:
  \url{https://www.cybrary.it/}
\BIBentrySTDinterwordspacing

\bibitem{paper-baciu}
\BIBentryALTinterwordspacing
O.-G. Baciu-Ureche, C.~Sleeman, W.~C. Moody, and S.~J. Matthews, ``The
  adventures of scriptkitty: Using the raspberry pi to teach adolescents about
  internet safety.''\hskip 1em plus 0.5em minus 0.4em\relax New York, NY, USA:
  Association for Computing Machinery, 2019. [Online]. Available:
  \url{https://doi.org/10.1145/3349266.3351399}
\BIBentrySTDinterwordspacing

\bibitem{tunc_claas_2015}
\BIBentryALTinterwordspacing
C.~Tunc and S.~Hariri, ``{CLaaS}: {Cybersecurity} {Lab} as a {Service},''
  \emph{Journal of Internet Services and Information Security (JISIS)}, Nov.
  2015. [Online]. Available:
  \url{http://isyou.info/jisis/vol5/no4/jisis-2015-vol5-no4-03.pdf}
\BIBentrySTDinterwordspacing

\bibitem{computer_science_education_research_group_at_the_university_of_canterbur_cryptographic_nodate}
\BIBentryALTinterwordspacing
C.~S. E. R.~G. at~the University~of Canterbury, ``Cryptographic protocols.''
  [Online]. Available:
  \url{https://classic.csunplugged.org/activities/cryptographic-protocols/}
\BIBentrySTDinterwordspacing

\bibitem{ncc_groups_cryptography_services_cryptopals_nodate}
\BIBentryALTinterwordspacing
N.~G.~C. Services, ``The {Cryptopals} {Crypto} {Challenges}.'' [Online].
  Available: \url{https://cryptopals.com/}
\BIBentrySTDinterwordspacing

\bibitem{beissinger_about_2021}
\BIBentryALTinterwordspacing
J.~Beissinger and B.~Saunders, ``About the {CryptoClub} {Project},'' 2021.
  [Online]. Available:
  \url{https://cryptoclubproject.uchicago.edu/about/about-the-project}
\BIBentrySTDinterwordspacing

\bibitem{mit_lincoln_laboratory_llcipher_2021}
\BIBentryALTinterwordspacing
{MIT Lincoln Laboratory}, ``{LLCipher},'' 2021. [Online]. Available:
  \url{https://www.ll.mit.edu/outreach/llcipher}
\BIBentrySTDinterwordspacing

\bibitem{lindmeier_keeping_2020}
\BIBentryALTinterwordspacing
A.~Lindmeier and A.~Mühling, ``Keeping secrets: {K}-12 students' understanding
  of cryptography,'' in \emph{Proceedings of the 15th {Workshop} on {Primary}
  and {Secondary} {Computing} {Education}}, ser. {WiPSCE} '20.\hskip 1em plus
  0.5em minus 0.4em\relax Association for Computing Machinery, 2020. [Online].
  Available: \url{http://doi.org/10.1145/3421590.3421630}
\BIBentrySTDinterwordspacing

\bibitem{gonzales20}
\BIBentryALTinterwordspacing
A.~I. Gonz\'{a}lez-Tablas, M.~I. Gonz\'{a}lez~Vasco, I.~Cascos, and
  A.~Planet~Palomino, ``Shuffle, cut, and learn: Crypto go, a card game for
  teaching cryptography,'' \emph{Mathematics}, 2020. [Online]. Available:
  \url{https://www.mdpi.com/2227-7390/8/11/1993}
\BIBentrySTDinterwordspacing

\bibitem{hutchison_teaching_2010}
\BIBentryALTinterwordspacing
L.~Keller, D.~Komm, G.~Serafini, A.~Sprock, and B.~Steffen, ``Teaching
  {Public}-{Key} {Cryptography} in {School},'' in \emph{Teaching {Fundamentals}
  {Concepts} of {Informatics}}.\hskip 1em plus 0.5em minus 0.4em\relax Springer
  Berlin Heidelberg, 2010. [Online]. Available:
  \url{http://link.springer.com/10.1007/978-3-642-11376-5_11}
\BIBentrySTDinterwordspacing

\bibitem{10.1145/3328778.3366816}
\BIBentryALTinterwordspacing
V.~\v{S}v\'{a}bensk\'{y}, J.~Vykopal, and P.~\v{C}eleda, ``What are
  cybersecurity education papers about? a systematic literature review of
  sigcse and iticse conferences,'' in \emph{Proceedings of the 51st ACM
  Technical Symposium on Computer Science Education}, ser. SIGCSE '20.\hskip
  1em plus 0.5em minus 0.4em\relax Association for Computing Machinery, 2020.
  [Online]. Available: \url{https://doi.org/10.1145/3328778.3366816}
\BIBentrySTDinterwordspacing

\bibitem{Younis2020}
\BIBentryALTinterwordspacing
Y.~A. Younis, K.~Kifayat, Q.~Shi, E.~Matthews, G.~Griffiths, and R.~Lambertse,
  ``Teaching cryptography using cypher (interactive cryptographic protocol
  teaching and learning),'' in \emph{Proceedings of the 6th International
  Conference on Engineering \& MIS 2020}, ser. ICEMIS'20.\hskip 1em plus 0.5em
  minus 0.4em\relax Association for Computing Machinery, 2020. [Online].
  Available: \url{https://doi.org/10.1145/3410352.3410742}
\BIBentrySTDinterwordspacing

\bibitem{CryptoTutor}
\BIBentryALTinterwordspacing
L.~Singleton, R.~Zhao, M.~Song, and H.~Siy, ``Cryptotutor: Teaching secure
  coding practices through misuse pattern detection,'' in \emph{Proceedings of
  the 21st Annual Conference on Information Technology Education}, ser. SIGITE
  '20.\hskip 1em plus 0.5em minus 0.4em\relax Association for Computing
  Machinery, 2020. [Online]. Available:
  \url{https://doi.org/10.1145/3368308.3415419}
\BIBentrySTDinterwordspacing

\bibitem{Chattopadhyay2019}
A.~Chattopadhyay, K.~Grondahl, J.~Ruckel, and T.~Everson, ``Secure coding and
  ethical hacking workshops with nao for engaging k-12 female students in cs,''
  in \emph{2019 Research on Equity and Sustained Participation in Engineering,
  Computing, and Technology}, 2019.

\bibitem{da_silva_criptolab_2018}
\BIBentryALTinterwordspacing
D.~J. G.~M. Da~Silva, G.~F. Guarda, and I.~F. Goulart, ``{CriptoLab}: {Um} game
  baseado em {Computação} {Desplugada} e {Criptografia},'' in \emph{Anais do
  {Workshop} sobre {Educação} em {Computação} ({WEI})}.\hskip 1em plus
  0.5em minus 0.4em\relax Sociedade Brasileira de Computação - SBC, 2018.
  [Online]. Available:
  \url{https://sol.sbc.org.br/index.php/wei/article/view/3483}
\BIBentrySTDinterwordspacing

\bibitem{scratch_foundation_scratch_nodate-1}
\BIBentryALTinterwordspacing
{Scratch Foundation}, ``Scratch - {Scratch} {Offline} {Editor}.'' [Online].
  Available: \url{https://scratch.mit.edu/}
\BIBentrySTDinterwordspacing

\bibitem{scratch_foundation_scratch_2020}
\BIBentryALTinterwordspacing
------, ``Scratch - {Annual} {Report},'' 2020. [Online]. Available:
  \url{https://scratch.mit.edu/annual-report}
\BIBentrySTDinterwordspacing

\bibitem{jin_2018}
G.~Jin, M.~Tu, T.-H. Kim, J.~Heffron, and J.~White, ``Game based cybersecurity
  training for high school students,'' in \emph{Proceedings of the 49th ACM
  Technical Symposium on Computer Science Education}, ser. SIGCSE '18.\hskip
  1em plus 0.5em minus 0.4em\relax New York, NY, USA: Association for Computing
  Machinery, 2018.

\bibitem{info11020121}
\BIBentryALTinterwordspacing
H.~Alqahtani and M.~Kavakli-Thorne, ``Design and evaluation of an augmented
  reality game for cybersecurity awareness (cybar),'' \emph{Information}, 2020.
  [Online]. Available: \url{https://www.mdpi.com/2078-2489/11/2/121}
\BIBentrySTDinterwordspacing

\bibitem{10.1145/3287324.3287450}
\BIBentryALTinterwordspacing
A.~L\'{e}deczi, M.~Mar\'{O}ti, H.~Zare, B.~Yett, N.~Hutchins, B.~Broll,
  P.~V\"{o}lgyesi, M.~B. Smith, T.~Darrah, M.~Metelko, X.~Koutsoukos, and
  G.~Biswas, ``Teaching cybersecurity with networked robots,'' in
  \emph{Proceedings of the 50th ACM Technical Symposium on Computer Science
  Education}.\hskip 1em plus 0.5em minus 0.4em\relax Association for Computing
  Machinery, 2019. [Online]. Available:
  \url{https://doi.org/10.1145/3287324.3287450}
\BIBentrySTDinterwordspacing

\bibitem{10.1145/3328778.3366878}
\BIBentryALTinterwordspacing
B.~Yett, N.~Hutchins, G.~Stein, H.~Zare, C.~Snyder, G.~Biswas, M.~Metelko, and
  A.~L\'{e}deczi, ``A hands-on cybersecurity curriculum using a robotics
  platform,'' ser. SIGCSE '20.\hskip 1em plus 0.5em minus 0.4em\relax
  Association for Computing Machinery, 2020. [Online]. Available:
  \url{https://doi.org/10.1145/3328778.3366878}
\BIBentrySTDinterwordspacing

\bibitem{10.1145/3282844}
\BIBentryALTinterwordspacing
H.~Hosseini, M.~Hartt, and M.~Mostafapour, ``Learning is child’s play:
  Game-based learning in computer science education,'' 2019. [Online].
  Available: \url{https://doi.org/10.1145/3282844}
\BIBentrySTDinterwordspacing

\bibitem{baciu-ureche_adventures_2019}
\BIBentryALTinterwordspacing
O.-G. Baciu-Ureche, C.~Sleeman, W.~C. Moody, and S.~J. Matthews, ``The
  {Adventures} of {ScriptKitty}: {Using} the {Raspberry} {Pi} to {Teach}
  {Adolescents} about {Internet} {Safety},'' in \emph{Proceedings of the 20th
  {Annual} {SIG} {Conference} on {Information} {Technology} {Education}}.\hskip
  1em plus 0.5em minus 0.4em\relax ACM, Sep. 2019. [Online]. Available:
  \url{https://dl.acm.org/doi/10.1145/3349266.3351399}
\BIBentrySTDinterwordspacing

\bibitem{9368440}
L.~M. Podila, J.~P. Bandreddi, J.~I. Campos, Q.~Niyaz, X.~Yang, A.~Trekles,
  C.~Czerniak, and A.~Y. Javaid, ``Practice-oriented smartphone security
  exercises for developing cybersecurity mindset in high school students,'' in
  \emph{2020 IEEE International Conference on Teaching, Assessment, and
  Learning for Engineering (TALE)}, 2020.

\bibitem{crypto_browserify}
\BIBentryALTinterwordspacing
{MIT}, ``{npm} {crypto-}{browserify},'' 2018. [Online]. Available:
  \url{https://www.npmjs.com/package/crypto-browserify}
\BIBentrySTDinterwordspacing

\bibitem{scratch_extensions}
\BIBentryALTinterwordspacing
------, ``{Scratch} {Extension}.'' [Online]. Available:
  \url{https://en.scratch-wiki.info/wiki/Scratch_Extension}
\BIBentrySTDinterwordspacing

\end{thebibliography}

\end{document}